\documentclass{article}
\usepackage{listings}
\usepackage{xcolor}
\usepackage{braket}
\usepackage{biblatex}
\usepackage{amsmath}
\usepackage{enumitem}
\usepackage{float}
\usepackage{graphicx}
\usepackage[affil-it]{authblk}
\usepackage[utf8]{inputenc}

\title{A Programming Language \\ for Quantum Oracle Construction}
\author{Ayush Tambde }
\date{October 2021}
\begin{document}
\maketitle
\definecolor{red}{RGB}{107,4,0}
\definecolor{blue}{RGB}{0, 43, 143}
\definecolor{green}{RGB}{0, 145, 75}
\lstset{
  language=Python,
  aboveskip=3mm,
  belowskip=3mm,
  showstringspaces=false,
  columns=flexible,
  basicstyle={\small\ttfamily},
  numberstyle=\tiny\color{gray},
  identifierstyle = \color{red},
  keywordstyle=\color{blue},
  morekeywords = {*,function, QuantumRegister, ClassicalRegister, QuantumCircuit,  oracle, super,elsif, measure},
  commentstyle=\color{green},
  moredelim = [s][\color{green}]{/**},
  rulecolor=\color{black},
  stringstyle=\color{red},
  breakatwhitespace=true,
  tabsize=3,
  frame=single
}

\section{Abstract}

Many quantum programs require circuits for addition, subtraction and logical operations. These circuits may be packaged within routines known as oracles. However, oracles can be tedious to code with current frameworks. To solve this problem the author developed Higher-Level Oracle Description Language (HODL) $-$ a C-style programming language for use on quantum computers $-$ to ease the creation of such circuits. The compiler translates high-level code written in \textit{HODL} and converts it into \textit{OpenQASM}[1], a gate-based quantum assembly language that runs on IBM Quantum Systems and compatible simulators. HODL is interoperable with IBM's QISKit framework.

\section{Introduction}

Quantum Computers were first conceptualized by the late physicist Richard Feynman[2] in a landmark paper in which he observed that classical techniques are inefficient when simulating quantum mechanics. Since the memory required to store a quantum state increases exponentially with the size of the system, Feynman proposed a new form of computer which would be capable of processing quantum information, hence the term \textit{Quantum Computer}, in which the fundamental unit of information is a two-level system known as a qubit. Quantum gates are applied to qubits in order to change their state[3]. 

Unfortunately current quantum frameworks make it tedious to construct classical functions in the form of quantum circuits. This was the primary motivation behind HODL where tasks such as addition, subtraction and multiplication on quantum states as well as relational operations can be performed in a simple yet expressive manner. This is aided by the HODL compiler's resource estimator which heuristically computes the total number of qubits required to run a program before run-time and therefore allows dynamic resizing of registers at compile-time. The compiler generates OpenQASM, meaning its output can be used with IBM's QISKit framework.

\section{The HODL Programming Language}

A standard HODL program consists of two main parts: 

\begin{itemize}
\item Declaration of data
\item Program statements to manipulate data
\end{itemize}

Statements and declarations are terminated with semicolons.\\

%flush to the left
\hspace*{-0.6cm} There are two types of data in HODL:

\begin{itemize}
\item \textit{Integer (\textbf{int})}
\item \textit{Quantum-Integer (\textbf{super})} 
\end{itemize}

These types distinguish between data allocated on two different devices, namely classical and quantum data. Storage for data is reserved using \textit{variable declarations}. Variables are represented by identifiers. The values of both data-types are expressed as integers. However, for quantum data, the integer $n$ represents the upper-bound of a uniform superposition of states. This provides a useful method for generating superposition states.
 
 Expressions in HODL can be composed of either classical or quantum data, and are constructed using a selection of operators (infix notation) similar to C. Operations are further discussed in Section \textbf{3.3}.
 
 \begin{itemize}
\item  Arithmetic operators are for addition, subtraction and multiplication (+, - , *, +=, -=, *=)
\item  Relational operators are for testing equality (==, !=) and order ($>$, $<$, $>=$, $<=$)
\item Boolean operators are for testing the value of Boolean functions (\&, $|$)
 \end{itemize}

\subsection{Functions and Oracles}

Subroutines are \textit{functions} that have the option to return a value or \textit{oracles} which, unlike functions, are first-class objects $-$ they return a memory address pointing to the location of the oracle in classical memory. Since all quantum operations are unitary and therefore reversible, the bodies of all subroutines in HODL are expanded inline.

Functions are declared using the keyword \textbf{function}. If a function returns a value, it is specified by preceding the keyword with the type.  This is followed by a function identifier followed by a list of parameters enclosed in parentheses. Finally, the function's body is enclosed in opening and closing braces. All parameters are passed-by-reference due to the no-cloning theorem which states that a quantum state cannot be cloned in its entirety [4].

The return statement saves whichever variable is returned from being uncomputed and the corresponding qubit register is returned and can be assigned a new identifier. The syntax for the return statement is the keyword \textbf{return}, followed by the variable name to be returned. Returning a variable is a classical instruction, and cannot be implemented inside quantum code blocks.

Although they have no return value, oracles are declared in a similar manner using the keyword \textbf{oracle} instead of the keyword \textbf{function}. The primary reason for oracles is the need to pass functions as parameters to other functions, this is because low-level memory manipulation is not supported and function-pointers cannot be used. Instead, oracles were introduced into the language for this purpose. When oracles are sent as input, a structure is passed which contains their address in memory amongst other metadata. Currently, oracles may only be passed to intrinsic functions such as \textbf{filter}, although it is a future development goal to allow oracles as parameters to user-defined subroutines. \\

\hspace*{-0.5cm}Intrinsic functions are \textbf{filter} and \textbf{mark}. 

The \textbf{mark} function must be within a quantum conditional. The function applies a phase of $\theta$ to all variables upon which the conditional control qubit is dependent on. The first parameter is an annotation for the programmer $-$ it specifies which variable the phase must be applied to and is only required so that one does not err by undesirably marking multiple variables. The second parameter specifies the value of the phase in terms of the keyword \textbf{pi}, since floating-point numbers are not currently supported. The function XORs the conditional control qubit with a qubit in the state $\frac{1}{\sqrt{2}}|0\rangle + e^{\theta}|1\rangle$.  

The \textbf{filter} function performs quantum search. It accepts as input a single oracle call, followed by a variable identifier representing the search space. The oracle must mark any states which are a solution to the search problem. Quantum search is further discussed in Section \textbf{4}.

\begin{figure}[H]
\caption{Functions and Oracles in HODL}
\begin{lstlisting}
# This is a comment

super function some_function(super foo, int bar) {
	# body
   # return some_var of type super
}

int function some_other_function() {
	# body
   # return some_var of type int
}

oracle some_oracle() {
	# body
}

function main() {
	# body
}


\end{lstlisting}
\end{figure}

\subsection{Type System}

Variable declarations in HODL begin with a type specifier, followed by a variable identifier. The type system is designed to indicate where a variable should be allocated. A variable of type \textbf{int} declares an integer on a classical computer, whereas a declaration introduced with the keyword \textbf{super} allocates qubits on a quantum computer. 

The keyword \textbf{super} for quantum variables provides a shorthand way to declare uniform superpositions. Values are assigned to either type of variable using the operator \textbf{=} as in C.

There is a limitation on the values that can be assigned to a quantum variable $-$ it must be a power of two, ie any quantum variable initialized with a value $x$ must satisfy $log_2(x) \in \textbf{N}$. This limitation enables the creation of a uniform superposition: $\sum_{i=0}^{x-1} \frac{1}{\sqrt{x}}|i\rangle$. When measured, this state will collapse into a basis state $|i\rangle$ with probability $(\frac{1}{\sqrt{x}})^2 = \frac{1}{x}$.  Reassignment of quantum variables is not allowed. \\

Measurement is performed using the keyword \textbf{measure}, followed by the variable name to be measured. Internally, the measurement operator creates a classical register referenced as ``creg\_var'' where ``var'' is the variable name. This is often the last step in many quantum algorithms, and at the moment HODL provides no mechanism for classical style post-processing.

\begin{figure}[H]
\caption{Type System in HODL}
\begin{lstlisting}
function main() {
	super a = 4;
   int b = 2;
}
\end{lstlisting}
\end{figure}

\subsection{Operations}
Operations can be performed on both quantum and classical data. Quantum variables may not be placed into a classical variable via any operation other than measurement. For example, the expression $c_2 = q + c_1$ where $q$ is quantum and both $c_i$ are classical, is not permitted, whereas the converse, $q_2 = c + q_1$ is. The compiler maintains a resource estimator at compile-time during which it tracks each operation and dynamically resizes quantum registers as required. This is particularly useful when performing arithmetic operations since they can alter the size of a register from $n$ qubits to $m$ qubits $-$ it saves the programmer from having to perform such tasks manually.

\subsection{Conditional Expressions}
HODL supports the if-else model that permeates throughout most modern programming languages. 

A conditional expression can be based on classical or quantum test conditions, but not both. Since quantum conditional expressions introduce entanglement, their bodies must purely be quantum.

The syntax for declaring an if-statement is to use the keyword \textbf{if} followed by a test condition enclosed in parentheses, succeeded by the conditional body enclosed within braces. 

The syntax is similar for an elsif statement, which can succeed either an if-statement or another elsif-statement. The only difference is instead of using the keyword \textbf{if}, one must use the keyword \textbf{elsif}. 

An else-statement signals the default case. It can succeed either an if-statement or an elsif-statement. To declare an else-statement, one must use the keyword \textbf{else}, followed by the body of the else-statement enclosed within braces.

\begin{figure}[H]
\caption{Conditionals in HODL}
\begin{lstlisting}
function main() {
	if(some_condition) {
    	# body
    }
    
   elsif(some_other_condition) {
    	# body
   }
    
   else {
    	# body
   }
}
\end{lstlisting}
\end{figure}

\subsection{Loops}
Loops are treated as classical constructs within HODL by expanding them inline during compile time. 
There are two forms of loops in the language:

\begin{itemize}
\item For loop
\item While loop
\end{itemize}

\subsubsection{For Loop}

The syntax for declaring a for loop is to use the keyword \textbf{for} followed by a series of three expressions enclosed in parentheses and separated by commas. These expressions must be classical.

The first expression should declare and/or initialize the classical data to be used in the loop. The second should specify the halt condition, that is, the circumstances required for the loop to terminate. As in C, the third condition specifies the modifications to be made to data on each iteration. After these three conditions the loop body is specified in braces.
 
\subsubsection{While Loop}

The syntax for declaring a while loop is to use the keyword \textbf{while}  followed by a single classical condition enclosed in parentheses. The condition specifies the circumstances required for the loop to run. After this condition the loop body is specified in braces.

\begin{figure}[H]
\caption{Loops in HODL}
\begin{lstlisting}
function main() {
   # FOR 
	for(int i = 0; i < 5; i+=1) {
    	# body
    }
    
   # WHILE
   while(cond) {
    	# body
    } 
}
\end{lstlisting}
\end{figure}

\subsection{Assembly Instructions}

HODL supports the basic quantum assembly instructions shown in Figure 5.

\begin{figure}[H]
\caption{Quantum Gates in HODL}
\begin{lstlisting}
function main() {
	# Hadamard gate
	H(foo);
   # X/NOT gate
   X(foo);
   # Y gate
   Y(foo);
   # Z gate
   Z(foo);
   # Rotate X gate
   RX(foo, angle);
   # Rotate Z gate
   RZ(foo, angle);
   # Rotate Y gate
   RY(foo, angle);
   # Apply phase
   P(foo, angle);
   # S gate
   S(foo);
   # T gate
   T(foo);
   # Controlled-Not gate
   CX(foo, bar);
   # Controlled-Z gate
   CZ(foo, bar);
   # Controlled phase
   CP(foo, bar, angle); 
}
\end{lstlisting}
\end{figure}

\subsection{Compiler Details}
The HODL compiler makes use of mechanisms such as register-size tracking in order to perform mathematical operations. Furthermore, ancillary registers used in such operations are dealt with internally and are uncomputed when not required and reset to zero automatically by maintaining an internal instruction tape intermediate program representation. They are referenced by the compiler as ancillaX where X is the number of ancillary registers in use at the time of creation decremented by one. Likewise, cmpX registers are used for storing results of relational operations. Classical expressions are evaluated at compile-time, leaving only quantum code to be compiled for later execution.

\begin{figure}
\caption{HODL Compiler Structure Flowchart}
\includegraphics[scale=0.6]{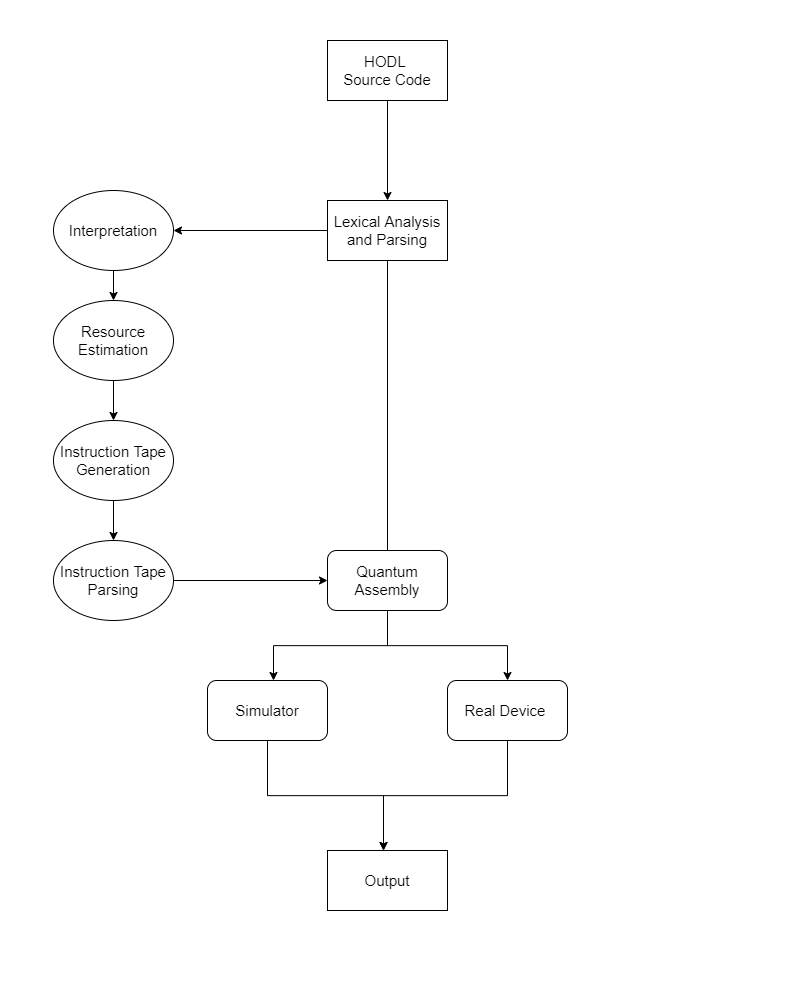}
\end{figure}
  \newpage

\subsubsection{Addition and Subtraction}

The addition operator in HODL is based on the Quantum Fourier Transform (QFT) [3] and is implemented as an optimized version of the method proposed by Draper[5].
The algorithm to perform $r$ = $x$ $+$ $y$ proceeds as follows:
\begin{enumerate}
\item Initialize two registers, $x$ and $y$ 
\item Initialize a register $r$ to $\ket{0}^{\otimes n}$, where $n$ = $floor(log_2(x+y)) + 1$
\item Apply $H^{\otimes n}$ to $r$ to obtain the state:\[(\frac{\ket{0} + \ket{1}}{\sqrt{2}})^{\otimes{n}} = \frac{1}{\sqrt{2^n}}(\ket{0} + \ket{1}) \otimes (\ket{0} + \ket{1}) \otimes  ... (\ket{0} + \ket{1})\]
\item Apply a series of controlled phase operations to store the Fourier Transform of $x$ in $r$:  \[\frac{1}{{\sqrt{2^n}}}  (\ket{0} + e^{2 \pi i 0.x_n}\ket{1}) \otimes
	(\ket{0} + e^{2 \pi i 0.x_{n-1}...x_n}\ket{1})  \otimes . . . (\ket{0} + e^{2 \pi i 0.x_1x_2...x_n}\ket{1})\]
\item Apply a series of controlled phase operations to add the Fourier Transform of $y$ into $r$.  $r$  now holds the sum of $x$ and $y$ in the Fourier Basis.\\ $\frac{1}{{\sqrt{2^n}}}  (\ket{0} + e^{2 \pi i (0.x_n + 0.y_n)}\ket{1}) \otimes
	(\ket{0} + e^{2 \pi i (0.x_{n-1}...x_n + 0.y_{n-1}...y_n)}\ket{1})  \otimes . . . (\ket{0} + e^{2 \pi i (0.x_1x_2...x_n + 0.y_1y_2...y_n)}\ket{1})$ \\
	
	$=$ $QFT|x + y\rangle$
	
\item Apply $QFT^\dagger$ to $result$ to retrieve $\ket{x + y}$  in the computational basis: \\ $QFT^\dagger QFT\ket{x + y} = \ket{x + y}$\\

\end{enumerate}
Note: To perform subtraction, the controlled operations in step 5 are inverted

\subsubsection{Multiplication}

The multiplication operator in HODL is similarly based on the QFT.
The algorithm to perform $j$ = $k$ $\cdot$ $l$ proceeds as follows:

\begin{enumerate}
\item Initialize registers $k$ as multiplicand, $l$ as multiplier and $j$ as $|0\rangle^n$ where n is the size of the output in bits and defaults to $n = size(k) + size(l) $ 
\item Apply $H^{\otimes n} $ to $j$, resulting in the state:

\[
(\frac{\ket{0} + \ket{1}}{\sqrt{2}})^{\otimes{n}})=  \frac{1}{\sqrt{2^n}}(\ket{0} + \ket{1}) \otimes (\ket{0} + \ket{1}) \otimes  ... (\ket{0} + \ket{1}) \] The goal is to transform the state described above to the Fourier Transform of $k\cdot l = j$  
\item  QFT$|j\rangle$ = $\frac{1}{\sqrt{2^n}}(\ket{0} + e^{2\pi i 0.j_1j_2...j_n} \ket{1}) \otimes (\ket{0} + e^{2\pi i 0.j_2...j_n}\ket{1}) \otimes  ... (\ket{0} + e^{2\pi i 0.j_n}\ket{1}) $, therefore it is desirable to obtain the relative phase factor $e^{2\pi i 0.j_1j_2...j_n}$

This can be achieved through multiplying the binary fractional forms of the multiplier and multiplicand: $(0.k) (0.l) = 0.kl = (\frac{k_1}{2^1} + \frac{k_2}{2^2}...\frac{k_n}{2^n}) \cdot (\frac{l_1}{2^1} + \frac{l_2}{2^2}...\frac{l_n}{2^n}) = $  \\

$\frac{k_1l_1}{2^2}$ + $\frac{k_1l_2}{2^3}$ ... $\frac{k_1l_n}{2^{n+1}}$  +

    $\frac{k_2l_1}{2^3}$ + $\frac{k_2l_2}{2^4}$ ... $\frac{k_2l_n}{2^{n+2}}$  +

            $\frac{k_n l_1}{2^{n+1}}$ + $\frac{k_n l_2}{2^{n+2}}$ ... $\frac{k_n l_n}{2^{2n}}$ 
            
            = $C$

\small{This is implemented as multi-controlled phase rotations ($P$) applied to a single qubit of $j$, to produce the phase $e^{2\pi i 0.j_1j_2...j_n} = e^{2\pi iC} $, where $P$ corresponds to the following matrix:}

  $\large{
  \begin{bmatrix}
1 & 0\\
0 & e^{2 \pi i \theta} 
\end{bmatrix}
}
 $
 \small
and $\theta$ is equal to the $\frac{1}{2^{x}}$ phase angles
   
\item Apply relative phases $e^{2\pi i 2^{m}C}$to each qubit $j_m$ and $0 \leq m < size(j) $ 

\item Apply $ QFT^{\dagger}$ (Inverse QFT) on $j$ to retrieve the product in the computational basis
\end{enumerate}

\section{Examples}

\subsection{Quantum Search}

The quantum search algorithm, first published in 1996 by Lov Grover [6], offers a polynomial-time advantage over the best known classical algorithm for searching for an item in an unordered dataset.\\

The algorithm is as follows:
\begin{enumerate}
\item   Initialize a superposition of the search space in state, $|s\rangle$.
\item  Apply an oracle, $O$, to $|s\rangle$. The oracle is designed in such a way that it applies a phase of $\pi$ if a term in the superposition fits a specified search constraint.
\item  Apply the diffusion operator ($2|s\rangle \langle s| - I$).
\item  Repeat steps 2 and 3, $(\frac{\pi}{4} \sqrt{\frac{N}{M}})$ times where $N$ is the size of the search space and $M$ is the number of solutions.
\item Measure state.
\end{enumerate}

The following is an application of the quantum search algorithm to search for all $x$ where $4x < 4$ and $0 \leq x \leq 7$.
The only solution is the state $0$, and the algorithm discovers it with high probability. 
The first figure depicts the algorithm written in HODL, and the second figure shows the same algorithm written in IBM's QISKit. 

\begin{figure}[H]
\caption{Example HODL program for Quantum Search Algorithm}
\begin{lstlisting}
# oracle takes a superposition as input
oracle some_oracle(super var) { 
	# if condition satisfied apply phase of pi to var
	if(var * 4 < 4) {
		mark(var,pi); 
	}
}

function main() {
	#declare a uniform superposition of 3 qubits
	super variable = 8;
	# "filter" function is an intrinsic function corresponding 
   # to the diffusion operator, accepting an oracle,
   # followed by the variable to apply the operator to
   filter(some_oracle(variable), variable);
	measure variable;
}

\end{lstlisting}
\end{figure}

\begin{figure}[H]
\caption{QISKit Code for Quantum Search Algorithm}
\begin{lstlisting}
# import libraries
from qiskit.circuit.library.arithmetic import IntegerComparator
from qiskit.circuit.library import QFT, GroverOperator
from qiskit.visualization import plot_histogram
from qiskit import *
from math import pi
# create and initialize registers
input_reg = QuantumRegister(3,name="input")
output_reg = QuantumRegister(5, name="output")
qc1 = QuantumCircuit(input_reg, output_reg)
qc1.h(input_reg)
qc = QuantumCircuit(input_reg, output_reg
# set up multiplication circuit based on QFT
phase = pi*1/2**2*4
phase_copy = phase*2
for i in range(5):
    qc.h(output_reg[i])
    for j in range(3):
        if i >= 3 and j == 0 or (j==1 and i==4):
            phase /= (2**1)
            continue
        qc.cp(phase,input_reg[j], output_reg[i])
        phase /= (2 ** 1)
    phase = phase_copy
    phase_copy *= 2
# apply inverse QFT
qft_circ = QFT(num_qubits=5).inverse()
qc.compose(qft_circ, qubits=[3,4,5,6,7], inplace=True)
qc1.compose(qc, qubits=range(8), inplace=True)
# compare with 4 (less than)
comparison_circ = IntegerComparator(5, 4, geq=False,name=
"comparison_circ")
circ_final = QuantumCircuit(15,3)
circ_final.compose(qc1, inplace=True)
circ_final.compose(comparison_circ, qubits=range(3,13), inplace=True)
# apply phase
circ_final.z(8)
# uncomputation
circ_final.compose(comparison_circ.inverse().to_gate(), 
qubits=range(3,13), inplace=True)
circ_final.compose(qc.inverse().to_gate(), qubits=range(8), inplace=True)
# diffusion operator applied once 
circ_final.h(range(3))
circ_final.x(range(3))
circ_final.mct([0,1,2], 13, 14, mode="basic")
circ_final.x(range(3))
circ_final.h(range(3))
# measurement
circ_final.measure(range(3), range(3))
\end{lstlisting}
\end{figure}
\subsection{Deutsch-Jozsa Algorithm}

The algorithm was first proposed by Deutsch [7] in 1985 and later revised in 1992 by Deutsch and Jozsa. 
Given a function $f$, this algorithm checks if it is constant or balanced. Note: $f$ is guaranteed to be one or the other $-$ if constant, $f$ returns either $0$ or $1$ for all inputs, otherwise it is balanced and returns $0$ for half of all inputs and $1$ for the other half. 

The algorithm proceeds as follows:
\begin{enumerate}
\item  Initialize state $|s\rangle$ as a superposition: $\frac{1}{\sqrt{2^n}}\sum_{i=0}^{2^n-1} |i\rangle$.
\item  Apply $f$ to $|s\rangle$ and XOR the result with a qubit in the $|-\rangle$ state.
\item  Apply Hadamard Gate ($H$), on $|s\rangle$.
\item  Measure $|s\rangle$.
\item  If $|s\rangle$ is measured to be zero, then $f$ is constant else it is balanced.
\end{enumerate}

The following is an application of the Deutsch-Jozsa algorithm for a function $f$ whereby $f(x) = 1$  if $ x + 7 > 14$ and $f(x) = 0$ if $x + 7 \leq 14$ where $0 \leq x \leq 15$. This function is balanced and must return a non-zero integer $-$ in this case the bit-string $1000_2$.

\begin{figure}[H]
\caption{Example HODL program for the Deutsch-Jozsa Algorithm}
\begin{lstlisting}
function deutsch_josza(super inputs) {
	# if condition satisfied store result in
	# 1-qubit register (called cmp0 internally)
	if(inputs + 7 > 14) {
		# XOR the contents of cmp0 with a qubit in the state |->
		mark(inputs,pi);
	}
}

function main() {
	# initialize superposition of 4 qubits
	super test = 16;
	deutsch_josza(test);
	# apply interference with Hadamard Gate
   H(test);
	# store result in a classical register creg_test
	measure test;	
}

\end{lstlisting}
\end{figure}

\begin{figure}[H]
\caption{QISKit Code for Deutsch Josza Algorithm}
\begin{lstlisting}
# import libraries for integer comparisons and addition
from qiskit.circuit.library.arithmetic import WeightedAdder,
IntegerComparator
from qiskit import QuantumRegister, ClassicalRegister, QuantumCircuit
# create circuits for addition and integer comparison
addition_circuit = WeightedAdder(7, [8,4,2,1, 4,2,1], name="adder_circ")
comparison_circuit = IntegerComparator(5, 15, name="comparison_circ")
# create registers and circuits for input and the integer "seven"
input_register = QuantumRegister(4, name="input_register")
seven = QuantumRegister(3, name="seven")
input_reg_to_circ = QuantumCircuit(input_register)
integer_seven_circ = QuantumCircuit(seven)
# apply HADAMARD on input to generate uniform superposition
input_reg_to_circ.h(input_register)
# flip all 3 bits in register to represent "7" in binary 
integer_seven_circ.x(seven)
# add summands as input to addition circuit
addition_circuit.compose(qc.to_gate(), qubits=[0,1,2,3], front=True, 
inplace=True)
addition_circuit.compose(qc1.to_gate(), qubits=[4,5,6], front=True,
inplace=True)
# create final circuit to hold all circuits
circuit_final = QuantumCircuit(22,4)
# append addition circuit in front of empty circuit
circuit_final.compose(addition_circuit.to_gate(), qubits=range(17), 
front=True, inplace=True)
# add comparison circuit with addition circuit
# with the input being the sum from the previous circuit
circuit_final.compose(comparison_circuit.to_gate(), 
qubits=[7,8,9,10,11,17,18,19,20,21], inplace=True)
# apply phase
circuit_final.z(17)
# uncomputation of circuits
circuit_final.compose(comparison_circuit.inverse().to_gate(), 
qubits=[7,8,9,10,11,17,18,19,20,21], inplace=True)
circuit_final.compose(addition_circuit.inverse().to_gate(), 
qubits=range(17), inplace=True)
# measure input register
circuit_final.measure(range(4), range(4))
\end{lstlisting}
\end{figure}

\section{Summary}

The author developed a new programming language for quantum computation that allows a higher-level description of oracle functions than available in existing frameworks. Although the language compiler can be used as a standalone tool, it was designed to be used alongside other OpenQASM-based frameworks such as QISKit. The author has open-sourced the language on GitHub[8].

\section{Acknowledgements}
The author thanks Prof David Abrahamson (Trinity College Dublin) for his guidance and supervision in the creation of this paper, and  Prof Brendan Tangney (Trinity College Dublin) and Dr Keith Quille (Technological University Dublin, and CSInc) for their insights and assistance. Thanks are also due to Dr Lee O'Riordan (Irish Center For High-End Computing), Dr Steve Campbell (UCD) and Dr Peter Rohde (UTS Australia) for their suggestions and feedback.

\section{Appendices}

\subsection{Lexical Specification in Regex}

digit = [``0"..``9"] \\
number = digit+\\
letter = [``a"..``z"] $|$ [``A"..``Z"]\\
identifier = letter (letter $|$ digit $|$ ``\_'')*\\
keyword = ``else''  $|$ ``elsif'' $|$ ``for'' $|$ ``function'' $|$ ``if'' $|$ ``int'' $|$ \newline
\hspace*{1.8cm}``measure'' $|$``oracle'' $|$ ``return'' $|$ ``super'' $|$ ``while'' \\
intrinsic\_function = ``mark'' $|$ ``filter'' \\
operator = ``+” $|$``-” $|$ ``*” $|$ ``+=” $|$ ``-=” $|$ ``*=” $|$ \\
\hspace*{1.8cm}``$<$” $|$ ``$>$” $|$ ``$<$=” $|$ ``$>$=” $|$ ``==” $|$ ``!=”\\

\hspace*{-0.6cm} Note: Keywords are reserved names and cannot be used as identifiers

\subsection{EBNF Specification of Syntax}

program = subroutine  \{[subroutine]\} \\
subroutine = (function $|$ oracle) \\
function = [type] \textbf{function} identifier parameters \textbf{\{} body [\textbf{return} identifier] \textbf{\}} \\
type = (\textbf{super} $|$ \textbf{int}) \\
parameters =  \textbf{(} [type identifier \{\textbf{,} [type identifier]\}] \textbf{)} \\
oracle = \textbf{oracle} identifier parameters  \textbf{\{} body \textbf{\}} \\
body = \{assignment $|$ fcall $|$ ocall $|$ operation $|$ loop $|$ cond\} \\
assignment = type identifier \textbf{=} (integer $|$ operation $|$ fcall)\\
fcall = identifier \textbf{(} \{[(type identifier)\ $|$ fcall $|$ ocall]\} \textbf{)} \textbf{;}\\
ocall = identifier \textbf{(} \{[(type identifier)\ $|$ fcall $|$ ocall]\} \textbf{)} \textbf{;} \\
operation = (identifier $|$ integer $|$ fcall $|$ operation)  operator \\
\hspace*{2cm}(identifier $|$ integer $|$ fcall $|$ operation)\\
loop = (for $|$ while)\\
for = \textbf{for} \textbf{(} assignment semicolon operation semicolon operation \textbf{) \{}  body \textbf{\}}\\
while = \textbf{while (} operation \textbf{) \{} body \textbf{\}}\\
cond = (if $|$ elsif $|$ else)\\
if = \textbf{if (} operation \textbf{) \{} body \textbf{\}}\\
elsif = \textbf{elsif (} operation \textbf{) \{} body \textbf{\}}\\
else = \textbf{else} \textbf{\{} body \textbf{\}}\\

\section{References}

[1] Andrew W. Cross, Lev. S Bishop, John A. Smolin, Jay M. Gambetta, Open Quantum Assembly Language, 2017. arXiv:1707.03429 [quant-ph] 
\newline \newline
[2]  Richard P Feynman. Simulating physics with computers, 1981. 
International Journal of Theoretical Physics, 21(6/7)
\newline \newline
[3] Nielsen and Chuang. Quantum Computation and Quantum Information, 2000. 
Cambridge Press
\newline \newline
[4] Wootters, W., Zurek, W. A single quantum cannot be cloned. Nature 299, 802–803 (1982). https://doi.org/10.1038/299802a0
\newline \newline
[5] Thomas G. Draper. Addition on a Quantum Computer, 2000. 
arXiv:quant-ph/0008033  
\newline \newline
[6] Lov K. Grover. A fast quantum mechanical algorithm for database search, 1996. 	arXiv:quant-ph/9605043 \newline \newline
[7] D. Deutsch and R. Jozsa. Rapid solution of problems by quantum compuation. Proceedings of Royal Society of London, A439:553–558 (1992) \newline \newline
[8] Ayush Tambde. https://github.com/at2005/HODL

\end{document}